# Theory and Applications of Two-dimensional, Null-boundary, Nine-Neighborhood, Cellular Automata Linear rules

[1]PABITRA PAL CHOUDHURY
[2] BIRENDRA KUMAR NAYAK, [3] SUDHAKAR SAHOO, [4]SUNIL PANKAJ RATH

[1]Applied Statistics Unit, Indian Statistical Institute, Kolkata, 700108, INDIA
pabitrapalchoudhury@gmail.com
[2]P.G.Department of Mathematics, Utkal University, Bhubaneswar-751004
bknatuu@yahoo.co.uk
[3, 4]Department of CSEA, SIT, Silicon Hills, Patia, Bhubaneswar-751024
sudhakar.sahoo@gmail.com and sunil.rath@wipro.com

*Abstract:* - This paper deals with the theory and application of 2-Dimensional, nine-neighborhood, null-boundary, uniform as well as hybrid Cellular Automata (2D CA) linear rules in image processing. These rules are classified into nine groups depending upon the number of neighboring cells influences the cell under consideration. All the Uniform rules have been found to be rendering multiple copies of a given image depending on the groups to which they belong where as Hybrid rules are also shown to be characterizing the phenomena of zooming in, zooming out, thickening and thinning of a given image. Further, using hybrid CA rules a new searching algorithm is developed called Sweepers algorithm which is found to be applicable to simulate many inter disciplinary research areas like migration of organisms towards a single point destination, Single Attractor and Multiple Attractor Cellular Automata Theory, Pattern Classification and Clustering Problem, Image compression, Encryption and Decryption problems, Density Classification problem etc.

## 1 INTRODUCTION

Cellular Automaton (CA) has emerged as a very useful concept. It has made understanding of many occurrences in nature easier. Though the concept is almost five decades old with John von Neumann [4] proposing it, yet only in the last two decades, researchers of various fields became interested to use the concept. The semiconductor technology is moving towards the sub-micron era and the system designers try to embed complex functions from software domain to hardware blocks on the silicon floor. While doing this, within a feasible limit of the complexity, the designers are forced to look for simple, modular, cascadable and reusable building blocks for implementing various complex functions. In this context, Cellular Automata (CA) is the right candidate to fulfill all the above objectives. Further the ever-increasing need for faster computing necessitates adoption of parallel processing architectures. It has been demonstrated by various authors [1, 2] that the parallel processing architecture built around the CA machine (CAM) is highly suitable for a variety of applications.

Wolfram et al. [5] studied one-dimensional CA with the help of polynomial algebra. Pries et al. [8] studied one-dimensional CA exhibiting group properties based on a similar kind of polynomial algebra. Later, Das et al. [9] extended the characterization of one-dimensional CA with the help of matrix algebra. Packard et al. [10] reported some empirical studies on 2D CA depending on five neighborhoods CA. Chowdhury et al. [11] extended the theory of 1-D CA built around matrix algebra for characterizing 2-D CA. However, emphasis was laid on special class of additive 2D CA, known as restricted vertical neighborhood (RVN) CA. In this class of 2-D CA the vertical dependency of a site is restricted to either the sites on its top or bottom, but not both. Characterization and applications of some particular uniform and hybrid 2D CA linear rules are reported in [3, 6]. Olu Lafe [12] uses Cellular Automata Transforms (CAT) for Data compression and encryption. Later Ikenaga et al. [7] studied Real-time morphology processing using highly parallel 2-D Cellular Automata $CAM^2$ (Cellular Automata on Content Addressable Memory). C. D. Thomas in his thesis [13] studied the evolution of Cellular Automata for image processing. Maji et al. [14] studied the theory and application of additive 1-D Cellular Automata rules for Pattern classification using matrix theory and characteristic polynomial.

Using matrices this paper contributes some extra theory of 2-D CA linear rules and touches some of the applications reported by different authors just discussed above. This requires introduction of the 2-D nine-neighborhood CA and the rules by which the dependencies of a cell are governed. In Section 2, the application of such rules on problem matrix is demonstrated which forms the basis of image processing. Section 3, presents the study of 512 linear Boolean functions, their matrix construction, structure of those matrices and their properties both in uniform as well as in hybrid CA. In Section 4, the effect of these linear rules on a given image is discussed. Particularly these rules are used for image transformations like translation, multiplication of one image into several, zooming- in and zooming-out, thickening and thinning of images. While the translation and multiplication can be carried out on any arbitrary image, structured images are needed for others. Multiple copies of any arbitrary image, so available, find innumerable applications in real life situation. In Section 5 a new algorithm called sweepers algorithm is developed using hybrid CA rules, which are found to be applicable in many important interdisciplinary research areas. Throughout this paper, transformation, maps and rules are used interchangeably.

## 2  MATHEMATICAL MODEL FOR 2-D CA

In 2-D Nine Neighborhood CA the next state of a particular cell is affected by the current state of itself and eight cells in its nearest neighborhood also referred as Moore neighborhood as shown in Figure-1. Such dependencies are accounted by various rules. For the sake of simplicity, in this section we take into consideration only the linear rules, i.e. the rules, which can be realized by EX-OR operation only. A specific rule convention that is adopted here is as follows:

| 64 | 128 | 256 |
|----|-----|-----|
| 32 | 1   | 2   |
| 16 | 8   | 4   |

[Fig-1]

The central box represents the current cell (i.e. the cell being considered) and all other boxes represent the eight nearest neighbors of that cell. The number within each box represents the rule number characterizing the dependency of the current cell on that particular neighbor only. Rule 1 characterizes dependency of the central cell on itself alone whereas such dependency only on its top neighbor is characterized by rule 128, and so on. These nine rules are called fundamental rules. In case the cell has dependency on two or more neighboring cells, the rule number will be the arithmetic sum of the numbers of the relevant cells. For example, the 2D CA rule 171 (128+32+8+2+1) refers to the five-neighborhood dependency of the central cell on (top, left, bottom, right and itself). The number of such rules is $^9C_0 + {}^9C_1 + \ldots + {}^9C_9$ =512 which includes rule characterizing no dependency.

These rules are grouped into nine groups in the following manner. Group-N for N=1, 2…. 9, includes the rules that refer to the dependency of current cell on the N neighboring cells amongst top, left, bottom, right, top-left, top-right, bottom-left, bottom-right and itself. The above-mentioned rule 171 therefore belongs to group-5. Thus group 1 includes 1, 2, 4, 8, 16, 32, 64, 128, and 256. Group 2 includes 3, 5, 6, 9, 10, 12, 17, 18, 20, 24, 33, 34, 36, 40, 48, 65, 66, 68, 72, 80, 96, 129, 130, 132, 136, 144, 160, 192, 257, 258, 260, 264, 272, 288, 320 and 384. Similarly rules belonging to other groups can be obtained. It can be noted that number of 1's present in the binary sequence of a rule is same as its group number.

### 2.1 Uniform and hybrid CA

The application of linear rules mentioned in the previous section can be realized on a problem matrix, where every entry is either 0 or 1. It may be mentioned that instead of applying the same rule to each entry of the problem matrix, it is admissible to apply different rules to different entries at the same time. While the former characterizes the uniform CA the latter characterizes hybrid CA.

**Example 2.2. (For Uniform CA)**
In figure 2, Rule 170 (2 + 8 + 32 + 128) is applied uniformly to each cell of a problem matrix of order (3 x 4) with null boundary condition (extreme cells are connected with logic-0 states).

$$\begin{pmatrix} 0 & 0 & 1 & 0 \\ 1 & 1 & 1 & 0 \\ 1 & 0 & 1 & 1 \end{pmatrix} \xrightarrow{Rule\ 170} \begin{pmatrix} 1 & 0 & 1 & 1 \\ 0 & 0 & 1 & 0 \\ 1 & 1 & 0 & 1 \end{pmatrix}$$

[Fig-2: shows the cell under consideration, will change its state by adding the states of its 4 orthogonal neighboring cells. 0+1+1+1=1. Similarly, Rule 170 is applied to all other cells.]

**Example 2.3. (For hybrid CA)**
Here is an example of a hybrid CA in null boundary condition where 3 rules (Rule 2, Rule 3, and Rule 4) are applied in 3 different rows (1st, 2nd and 3rd rows respectively) in the above problem matrix.

$$\begin{pmatrix} 0 & 0 & 1 & 0 \\ 1 & 1 & 1 & 0 \\ 1 & 0 & 1 & 1 \end{pmatrix} \xrightarrow{Rule\ 2, Rule\ 3, Rule\ 4} \begin{pmatrix} 0 & 1 & 0 & 0 \\ 0 & 0 & 1 & 0 \\ 0 & 0 & 0 & 0 \end{pmatrix}$$

[Fig-3]

The 2D CA behavior can be analyzed with the help of an elegant mathematical model, where we use two fundamental matrices to obtain row and column dependencies of the cells. Let the two-dimensional binary information matrix be denoted as $X_t$ that represents the current state of a 2D CA configured with a specific rule. The next state of any cell will be obtained by EX-OR operation of the states of its relevant neighbors associated with the rule. The global transformation associated with different rules can be made effective with the following two fundamental matrices, referred to as $T_1$ and $T_2$ in the rest of the paper:

$$T_1 = \begin{pmatrix} 0 & 1 & 0 \\ 0 & 0 & 1 \\ 0 & 0 & 0 \end{pmatrix} \text{ and } T_2 = \begin{pmatrix} 0 & 0 & 0 \\ 1 & 0 & 0 \\ 0 & 1 & 0 \end{pmatrix}$$

[Fig-4]

## 2.2 Characterization of 2-D CA

For the convenience of analysis, we will convert each of the rules into a transformation matrix denoted by $T_R$. We want to look at the transformation matrix $T_R$ such that $T_R$ operating on the current CA states X (The binary matrix of dimension (m x n)) generates the next state $[X']_{m \times n}$. The convention follows as in [1], [8]:

**Lemma 2.1:** The equivalent one-dimensional map matrix for any rule R can be represented as a tri-diagonal binary matrix is as follows:

$$T_R = \begin{pmatrix} D & U & \cdots & \cdots & \cdots & 0 & 0 & 0 \\ L & D & U & \cdots & \cdots & 0 & 0 & 0 \\ 0 & L & D & U & \cdots & 0 & 0 & 0 \\ \cdots & \cdots & \ddots & \ddots & \ddots & \cdots & \cdots & \cdots \\ \cdots & \cdots & \cdots & \ddots & \ddots & \ddots & \cdots & \cdots \\ \cdots & \cdots & \cdots & \cdots & \ddots & \ddots & \ddots & \cdots \\ 0 & 0 & 0 & \cdots & \cdots & L & D & U \\ 0 & 0 & 0 & \cdots & \cdots & 0 & L & D \end{pmatrix}_{mn \times mn}$$

[Fig 5: Where D, L and U are one of the following matrices of the order of n x n: [0], [I], [$T_1$], [$T_2$], [I+$T_1$], [I+$T_2$], [S] and [I+S], where S is sum of $T_1$ and $T_2$.]

This rule matrix $T_R$ of dimension (mn x mn) when multiplied with a 1D column matrix of order (mn x 1) obtained from the binary 2D (m x n) matrix in a row major order as reported in [1], [2], [8].

## 3  RULE MATRICES AND THEIR PROPERTIES

### 3.1  512-Linear Boolean functions

Let L be the set of 512 linear Boolean rules. We can define the operation $\oplus$ on the set of linear Boolean rules as follows: Rule m $\oplus$ Rule n = Rule k, Where if the binary representations of m, n and k are $m_1m_2…m_9$, $n_1n_2…n_9$ and $k_1k_2…k_9$, then $k_i = m_i \oplus n_i$  for i =1, 2, …9

Clearly Rule m $\oplus$ Rule m = Rule 0 and Rule m $\oplus$ Rule 0 = Rule m = Rule 0 $\oplus$ Rule m

Thus under the operation $\oplus$, L is a commutative group where each member is its own inverse.

Any member L can be uniquely expressed as the sum of the fundamental linear rules as shown in the following example.

**Example 3.1:** The expression for Rule 13.
Step 1- $(13)_{10} = (000001101)_2$
Step 2- Here, $0^{th}$ bit, $2^{nd}$ bit and $3^{rd}$ bit contains only 1 and the place values are
Step 3-  1 x $2^0$ = 1
        1 x $2^2$ = 4
        1 x $2^3$ = 8

Step 4- Rule 13 = Rule 1 $\oplus$ Rule 4 $\oplus$ Rule 8

**Other examples:**
Rule 3 = Rule 2 $\oplus$ Rule 1
Rule 35 = Rule 32 $\oplus$ Rule 2 $\oplus$ Rule 1

**Theorem 3.1:** As every number has a unique binary representation such expression is unique.

### 3.2 Uniform CA rules and its characteristics

#### 3.2.1  Structure of Rule matrices in uniform CA

If the problem domain is a binary matrix of order (m x n) then the rule matrices are of order (mn x mn) by lemma 2.1. Let $M_i$ denote the Rule matrix for Rule i, for i =0, 1, 2, 3…511. So, $M_1$ = Rule matrix for Rule 1, $M_2$ = Rule matrix for Rule 2 and so on.

Following two theorems are illustrated with examples shows the patterns found in all 512 linear rule matrices.

**Theorem-3.2:** When the Boolean rules are represented by rule matrices, the matrices corresponding to the fundamental Boolean rules are found to be related to each other in the following manner.
  $M_1$ = identity matrix
  $(M_2)^T = M_{32}$ i.e. transpose of Rule 2 matrix is Rule 32 matrix.
Similarly,
  $(M_4)^T = M_{64}$
  $(M_8)^T = M_{128}$
  $(M_{16})^T = M_{256}$
And other rule matrices can be generated from these 5 basic rule matrices by the help of two operations one is transpose denoted by T and other is addition modulo 2 denoted by $\oplus$.

**Example 3.2:** If M = {$M_1$, $M_2$, $M_4$, $M_8$, $M_{16}$} be the set of 5 basic matrices then
$M_6 = M_2 \oplus M_4$
$M_{33} = M_1 \oplus M_{32} = M_1 \oplus (M_2)^T$
$M_{34} = M_2 \oplus M_{32} = M_2 \oplus (M_2)^T$
$M_{15} = M_1 \oplus M_2 + M_4 \oplus M_8$
$M_{34} = M_{32} \oplus M_{64} = (M_2)^T \oplus (M_4)^T$

Observing different types of matrices used in Lemma 2.1, we have constructed two binary sequences $S_1$ and $S_2$ are as follows:

$S_1$= <u>111…1</u> 0 <u>111…1</u> 0 ………..0<u>111…1,</u> this sequence contains (n-1) '1's and one '0', then (n-1) '1's and one '0'and finally end with (n-1) '1'.
And
$S_2$= 0 <u>111…1</u> 0 <u>111…1</u> ……….0 <u>111…1</u>0, the sequence starts with a '0' followed by (n-1) 1's and one '0', (n-1) '1's and at the end with a '0'.

Where, n = number of columns in the given problem matrix.

**Theorem-3.3:** The 5 basic matrices possess a particular structure as follows.
$M_1$ = main diagonal contains all '1's and all other elements are '0'.
$M_2$ = super diagonal elements are arranged in a sequence like $S_1$ and all other elements are '0'.
$M_4$ = $(n+1)^{th}$ diagonal elements are arranged in a sequence like $S_1$ and all other elements are '0'.
$M_8$ = $n^{th}$ diagonal contains all '1's and all others elements are '0'.
$M_{16}$ =$(n-1)^{th}$ diagonal elements are arranged in a sequence like $S_2$ and all other elements are '0'.

**Example 3.3:** Let a problem is of order (2 x 2), then its corresponding rule matrix is of order (4 x 4)

$$M_1 = \begin{pmatrix} 1 & 0 & 0 & 0 \\ 0 & 1 & 0 & 0 \\ 0 & 0 & 1 & 0 \\ 0 & 0 & 0 & 1 \end{pmatrix}_{4 \times 4}$$

$$M_2 = \begin{pmatrix} 0 & 1 & 0 & 0 \\ 0 & 0 & 1 & 0 \\ 0 & 0 & 0 & 1 \\ 0 & 0 & 0 & 0 \end{pmatrix}_{4 \times 4} \quad M_4 = \begin{pmatrix} 0 & 0 & 0 & 1 \\ 0 & 0 & 0 & 0 \\ 0 & 0 & 0 & 0 \\ 0 & 0 & 0 & 0 \end{pmatrix}_{4 \times 4}$$

$$M_8 = \begin{pmatrix} 0 & 0 & 1 & 0 \\ 0 & 0 & 0 & 1 \\ 0 & 0 & 0 & 0 \\ 0 & 0 & 0 & 0 \end{pmatrix}_{4 \times 4} \quad M_{16} = \begin{pmatrix} 0 & 1 & 0 & 0 \\ 0 & 0 & 1 & 0 \\ 0 & 0 & 0 & 1 \\ 0 & 0 & 0 & 0 \end{pmatrix}_{4 \times 4}$$

[Figure-6: shows 5 basic Rule matrices over the domain (2x2). Here, m=2 and n=2. In $M_4$ $(n+1)^{th}$ i.e.$3^{rd}$ diagonal elements are arranged in a sequence like $S_1$. Similarly other matrices are formed. Numbering of the diagonals are done starting from main diagonal which is also the $0^{th}$ diagonal and super diagonal is the $1^{st}$ diagonal, next to the super diagonal is the $2^{nd}$ diagonal and so on.]

The above two observations, mentioned as Theorem-2 and Theorem-3 help us for generating all 511 linear rule matrices, and this way of construction is much more simpler than that was mentioned in the earlier paper [1] and [2].

**Corollary 3.2.1:** Except $M_1$ other 4 basic matrices {$M_2$, $M_4$, $M_8$, $M_{16}$} are lower triangular matrices whose diagonal elements are all 0 and therefore by Theorem-3.2 four fundamental matrices {$M_{32}$, $M_{64}$, $M_{128}$, $M_{256}$} are upper- triangular matrices with diagonal elements are all also 0.

### 3.2.2 Non-singular Rule matrix
Since Rule matrices are necessarily square matrices, the non-singular rule matrix called reversible CA [14] deserves to be studied closely because a non-singular matrix must produce cyclic state transition diagram having enumerable applications in Pattern Classification, Clustering, Encryption and Decryption, Data Compression etc.

Rule 1 is reversible because $M_1$ is non-singular and the matrices combined with $M_1$ contains 1's in their diagonal elements, also becomes non-singular. So, rules formed using Rule 1 are also reversible. Out of 512 rule matrices, following 31 rules are always reversible for arbitrary input matrix (m x n).

**List of 31 non-singular matrices/ Reversible CA:**

$Group\ 1\ rules - 1$

$Group\ 2\ rules - 3, 5, 9, 17, 33, 65, 129, 257$

$Group\ 3\ rules - 7, 11, 19, 13, 21, 25, 97, 161, 289, 193, 321, 385$

$Group\ 4\ rules - 15, 23, 29, 27, 225, 353, 417, 449$

$Group\ 5\ rules - 31, 481$

## 3.2 Hybrid CA rules and its characteristics

### 3.3.1 Structure of Rule matrices in hybrid CA

A hybrid CA as discussed in section 2 is one where each cell has its own local rule, which may be different from the local rule of any other cell. It may be observed that the rule matrices of order (mn x mn) that correspond to the linear Boolean rules applied to the binary matrix with entries 0 or 1 of order (m x n) is 512 in number. But, total number of matrices of order (mn x mn) with entries 0 or 1, is $2^{mn^2}$. The question arises as to which rules $2^{mn^2-512}$ matrices correspond. To find answer to such question the concept of hybrid matrices is introduced in the following way. Applying Rule 3 to a problem of order (2 x 2) means, here this rule is applied to each element and hence, it is a function over 9 variables as shown below.

$$\begin{matrix} \mathbf{0} & \mathbf{0} & & \mathbf{0} & \mathbf{0} & \text{Rule 3} \\ \mathbf{0} \begin{bmatrix} a & b \end{bmatrix} \mathbf{0} & \xrightarrow{\text{applied to 'c'}} & \begin{bmatrix} * & * \\ c+d & * \end{bmatrix} \\ \mathbf{0} \begin{bmatrix} c & d \end{bmatrix} \mathbf{0} & & \\ \mathbf{0} & \mathbf{0} & & \mathbf{0} & \mathbf{0} & \end{matrix}$$

[Fig-7: Shows that Rule 3 is applied to the element c and the result is $c \oplus d$. Other outputs are not shown denoted by * and hence Rule 3: $(A_1 \text{ x } A_2 \text{ x} \ldots A_9) \longrightarrow \{0,1\}$.]

But, globally when Rule 3 is applied to the whole matrix of order (2 x 2) then, actually we are playing over the matrix of order (4 x 4)(two rows and two columns are added in the boundary) and hence in this case Rule 3 is a function over 16 variables.

That is Rule 3: $(A_1 \text{x} A_2 \text{x} \ldots \text{x} A_{16}) \longrightarrow (M)_{m \text{ x } n}$ In this case, we can find $2^{2^{16}}$ Boolean functions. Out of which only $2^{16}$ functions are linear and Rule 3 belongs to this $2^{16}$ linear function in the case of 16 variables and not belongs to $2^9$ linear functions over 9 variables.

Therefore, the global effect of any rule and the effect of any hybrid rule on a binary matrix of order (m x n) is a function over ((m+2) x (n+2)) variables. That is any

Hybrid Rule: $(A_1 \text{ x } A_2 \text{ x } \ldots A_{(m+2) \text{ x } (n+2)}) \longrightarrow \{0,1\}$ where, $2^{2^{(m+2)\times(n+2)}}$ functions are possible out of which $2^{(m+2)\times(n+2)}$ functions are linear.

**Definition 3.1 (Hybrid matrices)**
Matrices connected with hybrid CA Rules are referred as hybrid matrices.

We know the behavior of all the 512 linear Boolean rules and also its matrix construction. For example Rule matrix for Rule 2 can also be constructed in another way and this new technique will help us to study the behavior of all possible hybrid matrices involving hybrid CA or non-uniform CA.

Suppose for example Rule 2 is applied to a problem matrix of order (3 x 3) then: 1st cell change its state by looking the state of the 2nd cell. 2nd cell change its state by looking the state of the 3rd cell. 3rd cell change its state by looking at the cell, which is not present in the matrix hence, its state is '0' as we are dealing with null boundary condition. 4th cell change its state by looking the state of the 5th cell. 5th cell change its state by looking at the 6th cell. 6th cell change its state by looking at the cell, which is not present in the matrix hence, its state is '0'. 7th cell change its state by looking at the 6th cell. 8th cell change its state by looking at the 6th cell. 9th cell change its state by looking at the cell, which is not present in the matrix hence, its state is '0'. By the definition of Rule 2, just given here we can construct our rule matrix for Rule 2, which is an order of (9 x 9).

**Construction of Rule 2 Matrix:**
1st row of the (9 x 9) matrix can be constructed from the behavior of the 1st cell. As discussed earlier on application of Rule 2, the state of the 1st cell is affected by the state of the 2nd cell. So, in the 1st row of the rule matrix only the 2nd position is 1 and other elements in that row are 0.
In 2nd row of the (9 x 9) matrix only the 3rd position is 1 and others are 0. Similarly other rows are constructed and the entire rule matrix is shown graphically. In general, "ith cell of the problem matrix corresponds to ith row of the rule matrix and the position of 1 in ith row depends on the cell positions in the problem doai affecting to the ith cell."

**Graphically**

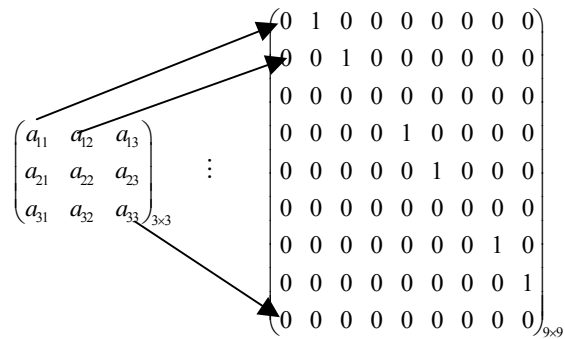

[Fig-8: Shows that ith cell of the problem matrix corresponds to ith row of the rule 2 matrix. The position of 1 in ith row depends on the cell positions in the problem domain that affects the cells by the property of Rule 2 just defined above.]

Here, Rule 2 is a uniform rule because all the cells behave exactly the same way (change occurs due to its right-hand side cell). But the construction rule discussed above is same for hybrid linear rules.

**Example 3.4:** Let us construct a rule matrix for a hybrid rule that behaves as follows:

**A hybrid rule:**
1. 1st cell change its state by looking at the value in 3rd position of the matrix.
2. 2nd cell change its state by looking at its bottom cell value i.e. 5th position.
3. 3rd cell change its state by looking at itself i.e. 3rd position in the matrix.
4. 4th cell change its state by looking at the last position i.e. 9th position.
5. 5th cell change its state by computing the sum of 1st, 2nd and 3rd position.
6. 6th cell change its state by looking at itself i.e.9th position in the matrix.
7. 7th cell change its state by computing the sum of the states in1st and 9th position.
8. 8th cell change its state by seeing its bottom position, which is 0.
9. Finally, Rule 170 is applied on 9th cell i.e. by computing the sum of the states at 8th position, 6th position, 0 and 0.

**Matrix Construction:**

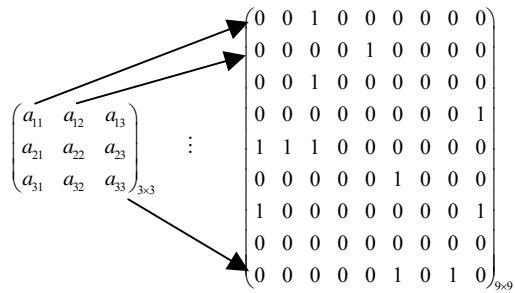

[Fig-9: Shows that i[th] cell of the problem matrix corresponds to i[th] row of the hybrid matrix. The position of 1 in i[th] row depends on the cell positions in the problem domain that affects the cells by the property of the corresponding hybrid rule.]

Reverse operation can also be possible that is given any arbitrary matrix of order (mn x mn), one should be able to tell how each cell of (m x n) getting changed. Here, i[th] cell depends on those positions of m x n matrix where, i[th] row of (mn x mn) matrix contains exactly the value 1.

In our previous sections, we have discussed the mathematical properties of all the linear rules, which are mostly relevant to our image processing work.

## 4 APPLICATION IN IMAGE PROCESSING

The 2D CA linear rules, which are used for image processing work reported in this paper, are 2, 4, 8,16,32,64,128,256 and these are of group 1 rules according to our classification shown in Section 2. Moreover, in the case of multiplication of images, 2D CA linear rules from different groups are used.

For applying 2D CA linear rules in image processing, we take a binary matrix of size (100x100). We map each element of the matrix to a unique pixel on the screen (Using BGI graphics of Turbo C++ 3.0) and we color a pixel White for 0 & Black for 1 for the matrix elements. This is the way, how the image is drawn within an area of (100x100) pixels, which we indicate by coloring its boundary by Red. We apply 2D CA linear rules on that (100x100) matrix (Both in regular and in some cases hybrid form) and each time the rule is applied using the changed matrix and a new image is redrawn. Initial image is taken by making some elements of the (100x100) matrix as 1, in a suitable fashion. We call this initial image as 'seed image'.

### 4.1 Translation of images

It may be observed that the fundamental rules namely rule 2, 4, 8, 16, 32, 64, 128, 256, which belong to group 1 cause translation of the image in the directions mentioned against each rule in the table-1. The extent of shift is dependent on the number of repetitions of the application of such rules as illustrated in the following figures (fig-8 and fig-9).

| Direction of the translation of the image | Rule of Group -1 to be applied |
|---|---|
| Top | 8 |
| Bottom | 128 |
| Left | 2 |
| Right | 32 |
| Top-Left | 4 |
| Top-Right | 16 |
| Bottom-Left | 256 |
| Bottom-Right | 64 |

Table-1

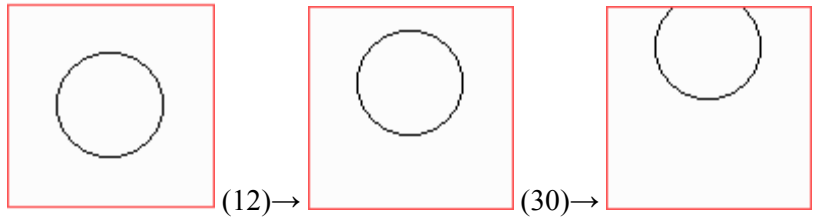
[Fig-10: Application Rule-8 (To a circle of radius - 26 pixels)]

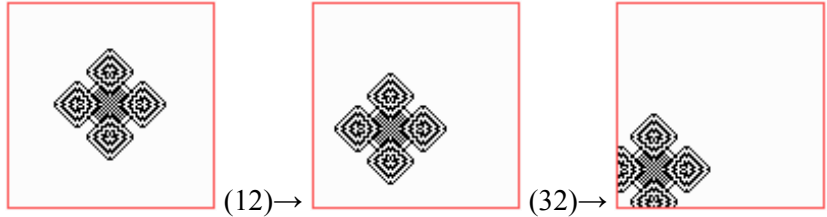
[Fig-11: Application of Rule – 256]

## 4.2 Multiple copies of any arbitrary image

The rules other than the fundamental rules create multiple copies of the given image, the number of copies being the same as the group number to which the applied rule belongs. Thus the maximum number of copies an image can have on application of such rules is 9, because the maximum group number is 9. It is observed that the multiple copies arise only when number of repetition is $2^n$ (n=1, 2, 3,…). The following figure illustrate the above assertion for n=5.

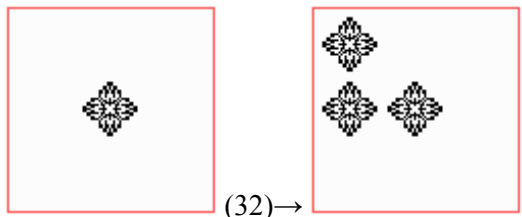
[Fig-12: Application Rule-7, Group-3]

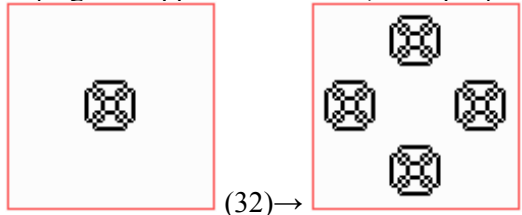
[Fig-13: Application Rule-170, Group-4]

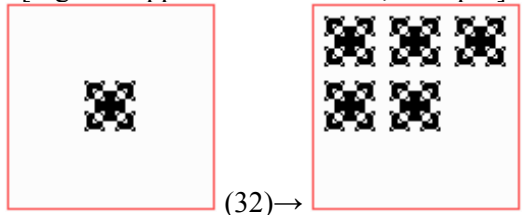
[Fig-14: Application Rule-31, Group – 5]

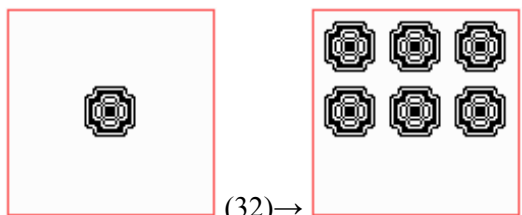
[Fig-15: Application Rule-63, Group – 6]

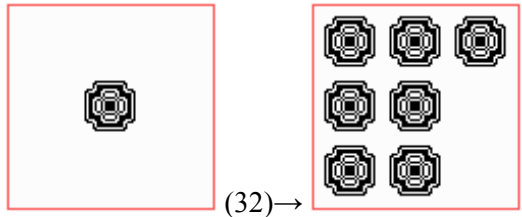
[Fig-16: Application Rule - 415, Group – 7]

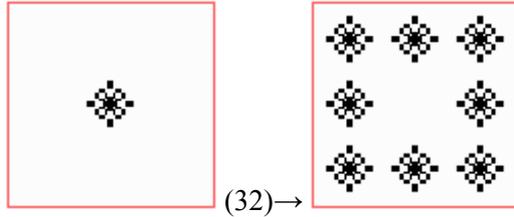
[Fig-17: Application Rule - 510, Group – 8]

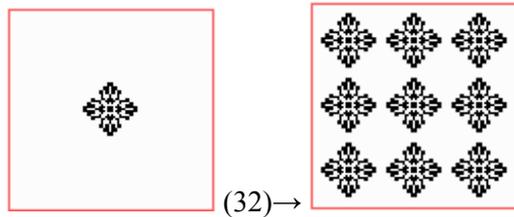
[Fig-18: Application Rule - 511, Group – 9]

It may also be observed that the rules belonging to same group though create equal number of copies, the distributions of these copies differ. For example after applying rule-5 of group-2 on an image (centered), the two copies of that image are found in center & upper left corner of the matrix (matrix refers to the bounded area, where images are displayed on the screen). But when rule- 68 of the same group is applied on the same image (centered), the two copies of original image are found in upper-left & lower-right corners of the matrix.

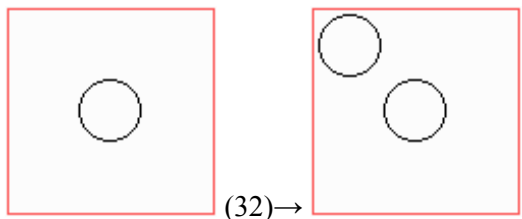
[Fig-19: Application Rule – 5, Group-2]

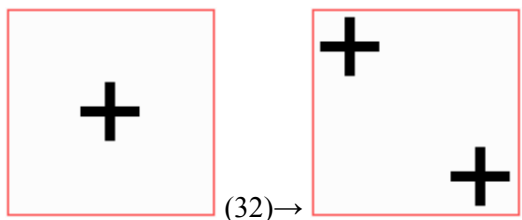
[Fig-20: Application Rule-68, Group-2]

Such copies of the image, it may however be mentioned, are formed within the display matrix provided that maximum length of the image considered in all directions, not exceeding 31% of the length of row or column (square matrix) of the display matrix (Here it is 100).

Further research effort is required to explain why only $2^n$ time's repetition of the rules causes the appearance of multiple copies of any arbitrary image.

## 4.3 Zooming in, zooming out and thickening, thinning of symmetric images

When the application of the above said rules are such that different rules are applied to different pixels of a symmetric image, the shape of the image changes; zooming in, zooming out, thickening and thinning of the image can be observed. The following figure illustrates these phenomena.

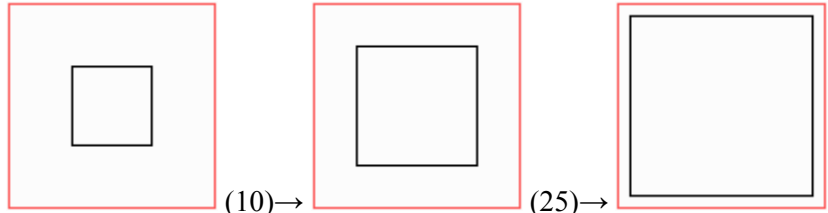

[Fig- 21: Application of Rules–2,32,8,128  (To a square of length - 40 pixels, transformation is zooming in)]

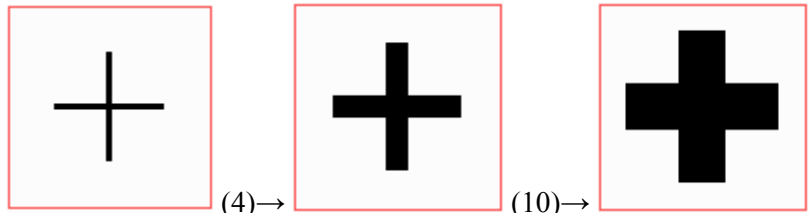

[Fig-22: Application of Rules – 2,32,8,128  (To a + sign of length - 55 pixels and breadth–3 pixels, transformation is zooming in)]

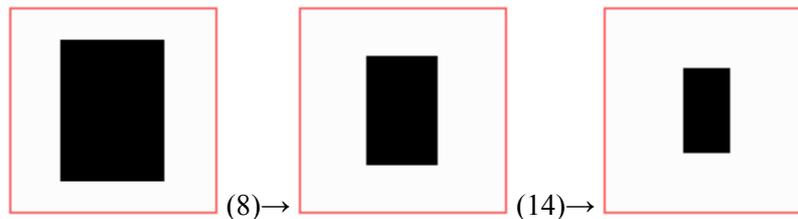

[Fig-23: Application of Rules-32, 2, 128,8  (To a rectangle of length - 50 & breadth – 70 pixels, transformation is zooming out)]

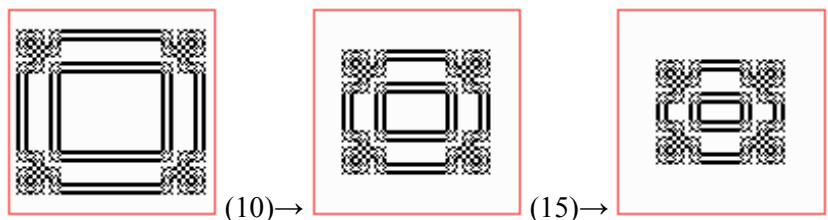

[Fig-24: Application of Rules-32, 2, 128,8  (To a complex pattern, transformation is zooming out)]

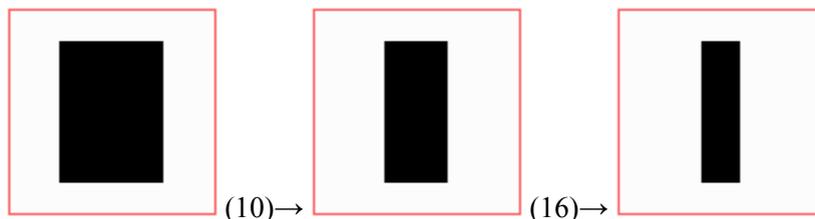

[Fig-25: Application of Rules-32, 2,1,1  (To a rectangle of length-50 & breadth-70 pixels, transformation is horizontal thinning)]

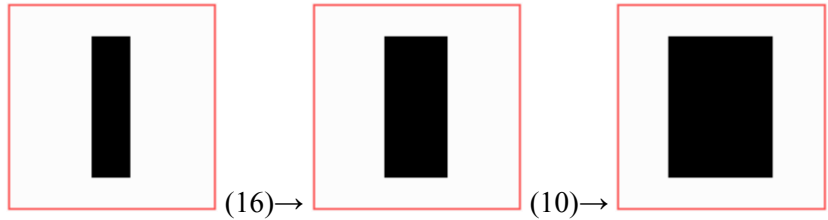

[Fig-26: Application of Rules – 2,32,1,1 (To a rectangle of length - 50 & breadth – 70 pixels, transformation is horizontally thickening, reverse of the transformations in Fig-19)]

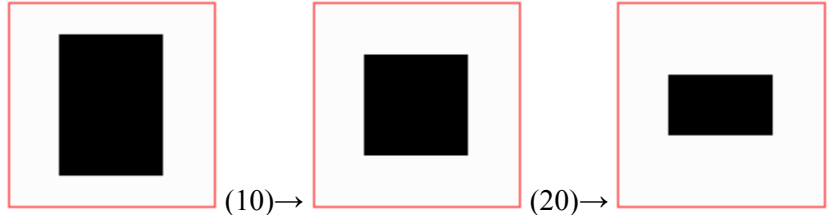

[Fig-27: Application of Rules – 1,1,128,8  (To a rectangle of length - 50 & breadth – 70 pixels, transformation is vertically thinning)]

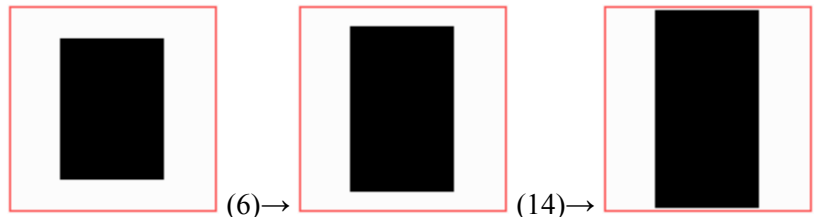

[Fig-28: Application of Rules–1, 1, 8, 128 (To a rectangle of length - 50 & breadth – 70 pixels, transformation is vertically thickening)]

In such application, the display matrix is partitioned into four regions as shown in fig-29.

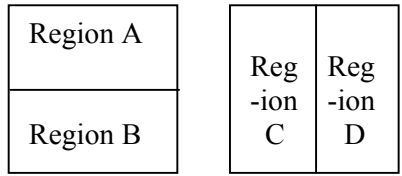

[Fig-29: Division of (100x100) 2 D CA region for hybrid transform]

   Two rules (may be different) are applied to two sections Region A and Region B followed by application of another two rules (may be different) to the sections Region C and Region D. These two applications called a single hybrid procedure, denoted as Hybrid (a, b, c, d), where a, b, c, d are four rules applied respectively to those four regions namely Region A, Region B, Region C and Region D. It may be observed that Hybrid (2,32,8,128), implements Zooming in (Fig-21 and Fig-22) where as Hybrid (32,2,128,8), implements Zooming out (Fig-23 and Fig-24). Hybrid (32,2,1,1), implements horizontal thinning where as Hybrid (2,32,1,1), implements horizontal thickening (Fig-25 and Fig-26). Hybrid (1,1,128,8), implements vertical thinning where as Hybrid (1,1,8,128), implements vertical thickening (Fig-27 and Fig-28).

## 5. SWEEPER'S ALGORITHM AND ITS APPLICATIONS
      In this section a searching algorithm is proposed using hybrid CA rules which is found to be applicable in many inter-disciplinary research areas. This technique is very similar to the way a sweeper sweeps haphazardly and puts it in one corner of a room that's why we call it as sweepers Algorithm. This tells a way to apply different CA rules in different regions by rotating a fixed axis about a fixed point as follows:

### 5.1 Sweeper's Algorithm
In this algorithm three basic requirements are:
1. An Axis that can divide the searching space into two areas where two different rules can be applied simultaneously.
2. A point on the axis where the axis can rotate.
3. An angle of rotation.

In a single iteration, the axis will rotate 'N' times with each time as a different hybrid CA rules. the value of N can be calculated using the formulae, N=360/2* (The angle of rotation)

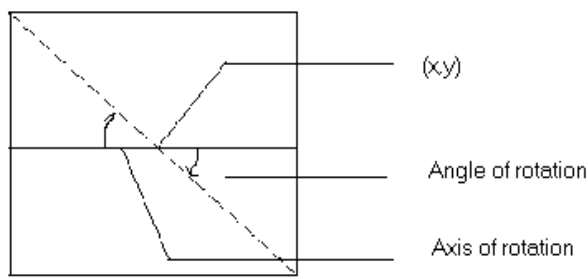

[Fig-30]

**Algorithm 5.1**
Step1:- A Monochromatic ".bmp" file is taken as the input where n is the number of iteration and the size of the matrix is (X x Y), the point of destination = (x, y), the angle of rotation = 45 (in degree), the number of rotations in one iteration, N is [360/2*45] = 4.
Step2- do the following for n times.
```
        {
          for ( i = 1 to x & j = 1 to Y)   // Two CA rules are applied with respect to Horizontal Axis
              Apply Rule 128
          for ( i = x+1 to X & j = 1 to Y)
              Apply Rule 8
          for ( i-j <= x-y)                // Two CA rules are applied with respect to Diagonal Axis ( \ )
              Apply Rule 16
          for ( i-j > x-y)
              Apply Rule 256
          for (i = 1 to X & j = 1 to y)    // Two CA rules are applied with respect to Vertical Axis
              Apply Rule 32
          for(i = 1 to X  & j = y +1 to Y)
              Apply Rule 2
          for ( i+j <= x+y)                // Two CA rules are applied with respect to Diagonal Axis ( / )
              Apply Rule 64
          for ( i+j > x+y)
              Apply Rule 4
        }
```
Step3- Store the resultant matrix in "Result.bmp" as the output.

Different CA rules are applied in various for loops in step 2 of the above algorithm. For example in the first and second inner for loops, CA rules 128 and 8 are applied on two different regions separated by a horizontal axis as shown in figure-31. Four each value of n in the outer loop four sequential steps have to be performed by dividing the regions by horizontal axis, diagonal axis, Vertical axis and reverse diagonal axis respectively as shown in fig-31. For boundary cells if the pixel values are 1, rules are never applied otherwise the 1's will get lost. This algorithm can be easily implemented using the VLSI architecture of Cellular Automata Machine (CAM) reported in [1, 2, 3, 6] and also using highly parallel 2-D Cellular Automaton architecture called CAM$^2$ (Cellular Automata on Content Addressable Memory) reported in [7]. CAM$^2$ performs one morphological operation for basic structuring elements within $30 \mu s$. This means that more than 1000 operations can be carried out on an entire 512 x 512 pixel image at video rates (33 ms). CAM$^2$ can also handle an extremely large and complex structuring element of 100 x 100 at video rates.

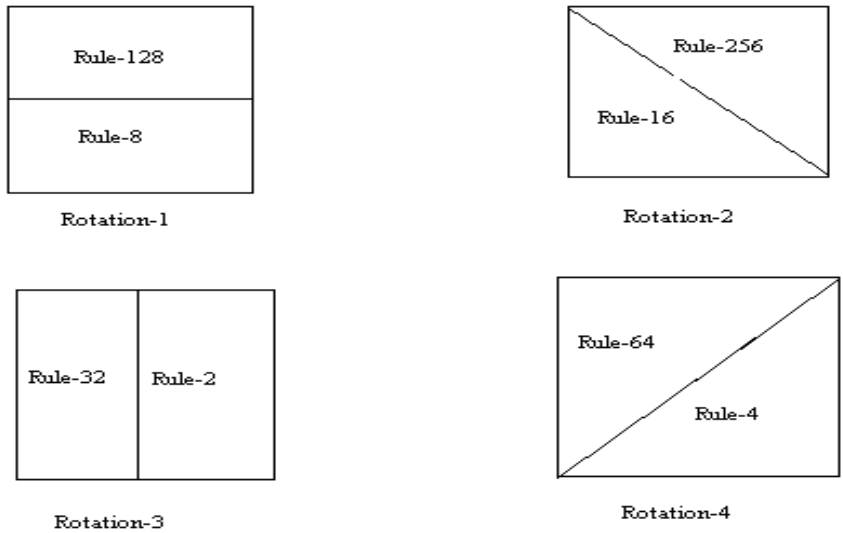

[Fig-31: Shows hybrid CA Rules are applied in a single iteration]

## 5.2 Applications of Sweeper's Algorithm

### 5.2.1 Migration of organisms towards a single point destination
Here sweeper's algorithm is used to show the different steps of migration of organisms in a particular environment towards a single destination. The organisms can be unicellular or multi-cellular. In case of unicellular organisms it is considered as the phenomenon of Chemo taxis as described below.

**Chemotaxis:** Chemotaxis is a migratory response [15] elicited by chemicals. Unicellular (e.g. protozoa) or multicellular (e.g. worms) organisms are targets of the substances. A concentration gradient of chemicals developed in a fluid phase guides the vectorial movement of responder cells or organisms. This is important for bacteria to find food (for example, glucose) by swimming towards the highest concentration of food molecules, or to flee from poisons (for example, phenol). In addition it has been recognized that mechanisms that allow chemotaxis in animals can be subverted during cancer metastasis. Fig-4 shows the behavior of eucariotic chemotaxis.

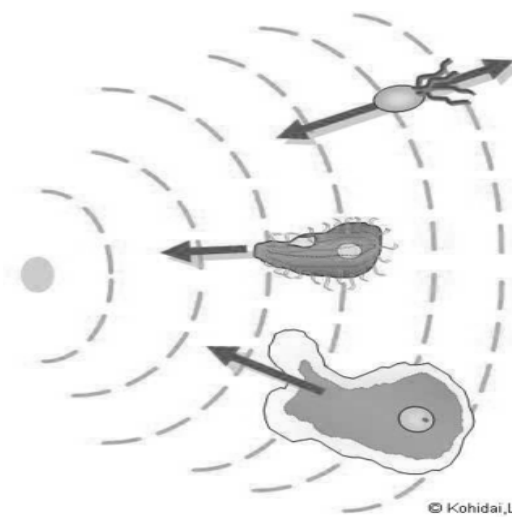

[Fig-32 Eucariotic Chemotaxis]

**5.2.2 Simulation Results**
Consider a uniformly or non-uniformly distributed environment (2-D space) of organisms.
The following assumptions are taken for simulation of chemotaxis in animals in MATLAB
1. The destination is a single point
2. All the individual organisms have the same property of being attracted towards the destination.
3. The size of the population remains constant throughout the course of motion.
4. The movement of the organisms is planar and no overlapping is allowed.

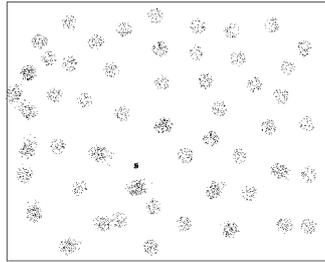

[Fig 33: Shows the initial position of organisms in a 2-D space of size (500 x 400) and the destination point is at (200,200)]

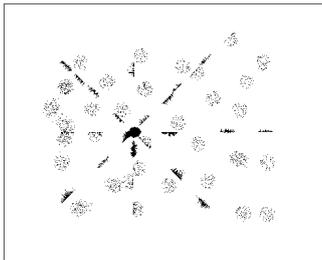

[Fig-34: After 20 iterations]

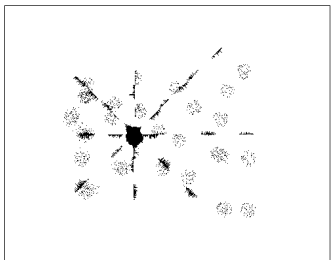

[Fig 35: After 30 iterations]

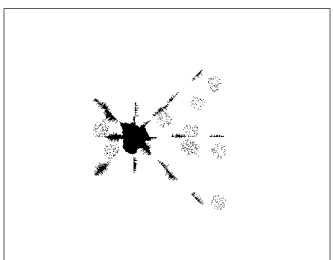

[Fig-36: After   45 iterations]

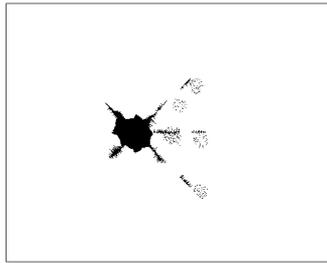
[Fig-37: After   55 iterations]

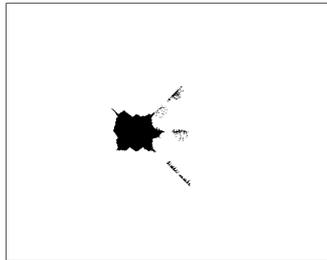
[Fig-38: After 65 iterations]

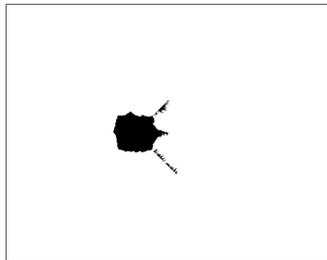
[Fig-39: After 75 iterations]

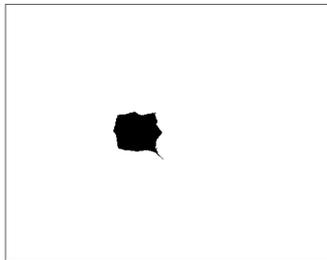
[Fig-40: After 90 iterations]

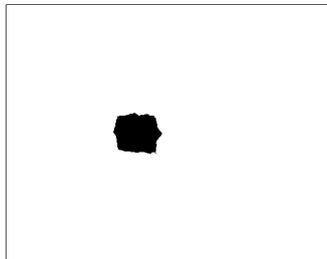
[Fig-41: After 100 iterations]

### 5.2.3 Other applications of Sweepers Algorithm
Except this biological application, sweepers algorithm can be applied in several other areas like  Informed search used in artificial intelligence,  Clustering problem,  text and image compression,  Cryptography,  finding single attractor and multiple attractors  Cellular Automata theory,  Pattern Classification, Density Classification problem etc. The detail application part can be dealt in separate papers.

# 6 CONCLUSION

In various purposes image processing and other morphological operations is needed, and there are various software and other means to do it. But here on using two-dimensional Cellular Automata rules various image processing functions can be easily embedded in the VLSI chip form using CAM or $CAM^2$. The result is, the time taken to process any image or any searching technique is the order of nanoseconds or microseconds respectively, which will be much faster compared to the other means of image processing and morphological operations. Although only some important primary image transformations are being reported for symmetric binary images, we feel that the work can be extended further for any other complex image transformations and also for asymmetric binary as well as color images. Our future research efforts are on towards that direction. Again the sweepers algorithm first time proposed in this paper using 2-D hybrid CA linear rules, spans many other inter-disciplinary research areas, some of them are reported and others to be explored in the future.